# Atomic Origin of Ti Deficient Dislocation in SrTiO$_3$ Bicrystal and Their Electronic Structures

Xujing Li[1,2], Shulin Chen[2,3], Mingqiang Li[2], Jingmin Zhang[2], Xiumei Ma[2], Kaihui Liu[4,5], Xuedong Bai[1,5*], Peng Gao[2,5,6*]

[1]Beijing National Laboratory for Condensed Matter Physics and Institute of Physics, Chinese Academy of Sciences, Beijing 100190, China

[2]Electron Microscopy Laboratory, School of Physics, Peking University, Beijing, 100871, China

[3]State Key Laboratory of Advanced Welding and Joining, Harbin Institute of Technology, Harbin 150001, China

[4]State Key Laboratory for Mesoscopic Physics, School of Physics, Peking University, Beijing, 100871, China

[5]Collaborative Innovation Centre of Quantum Matter, Beijing 100871, China.

[6]International Center for Quantum Materials, School of Physics, Peking University, Beijing, 100871, China

E-mails: p-gao@pku.edu.cn; xdbai@iphy.ac.cn.

**Abstract**

Dislocations in perovskite oxides have important impacts on their physical and chemical properties, which are determined by their unique atomic environments. In the present study, the structure of dislocations in a 10° low-angle grain boundary of SrTiO$_3$ (STO) is characterized by spherical aberration corrected scanning transmission electron microscopy (Cs-STEM) and spectroscopy. In contrast to previous studies, the deficiency instead of enrichment of titanium (Ti) is observed at the dislocation cores mainly due to the Sr substitution and under occupancy of Ti. The presence of oxygen vacancies and partially reduced Ti are also detected at the Ti deficient dislocations cores. These findings indicate the atomic structure of dislocations can be very different even they have the same Burges vectors. Controllable elemental segregation in the dislocations and grain boundaries via bicrystal engineering should be very useful for design of devices with novel functions.

**Keywords:** SrTiO$_3$ bicrystal, grain boundary, Ti deficiency, electronic structure

**Introduction**

Dislocations that ubiquitously exist in crystal materials and devices can be manipulated to realize peculiar physical or chemical properties[1-3], including unique electronic and ionic transportation properties[4,5], switchable resistance[6,7], ferroelectric polarization suppression and ferroelectric domain wall pinning[8,9], enhancement of current densities in superconducting conductors[10,11] and low-field magnetoresisitance in magnetic films[12]. Therefore, it's of great importance to discover the atomic structure and subsequently control the properties of the dislocations. Since low-angle bicrystal structure with grain boundary (GB) consists of regulable and periodic dislocation arrays, bicrystal engineering provides a very useful pathway to design dislocations with controllable structure and properties by changing tilt angle and elemental doping. Many efforts have been made to fabricate bicrystals and reveal their physical and chemical properties[13-19]. Among of them, $SrTiO_3$ (STO) bicrystals have been studied extensively as a model system for perovskite structure[20,21]. Generally, the fabrication of STO bicrystals experiences heat treatments under a uniaxial load in air [22,23]. In this circumstance, the dislocation cores in the grain boundary plane are usually found to be Ti rich[22,24]. In fact, in the electroceramic STO, the Ti rich dislocation cores leading to positively charge dislocations can explain many interesting physical phenomena on the electrical activities such as the presence of space charge layer[25] and enhanced ionic conductivities[26,27]. The formation of Ti enrichment is believed to originate from the generation of Sr vacancies at high temperature and tensional stress at the dislocation cores induced a transition of $TiO_6$ octahedrons from corner sharing to edge sharing[23,28,29], forming localized Ti-O rocksalt phase. However, such a model may be not applicable to all the STO dislocations because the atomic structure and chemistry of the dislocations should significantly rely on the history of sample treatment.

In this work, we demonstrate that the STO dislocation cores in a low angle bicrystal can be Ti deficiency instead of Ti enrichment by controlling the fabrication conditions. Such Ti deficient dislocations are obtained by diffusion bonding at high temperature in the atmosphere of $N_2$. The grain boundary consists of two alternative dislocations labeled as A (Ti-O plane terminated) and B (Sr-O plane terminated), which have the same Burges vector but different atomic arrangements and chemistry. From the quantitative energy dispersive X-ray spectroscopy (EDS) measurements, the ratio of Sr and Ti is 0.83:0.71 for the dislocation core A, and 0.88:0.67 for the core B compared to 1:1.02 in the bulk (ideally it should be 1:1 as a perovskite structure). Also, at the dislocation cores, the Ti is partially reduced from $Ti^{4+}$ to $Ti^{3+}$ due to the presence of oxygen vacancies, which has different electronic structures compared with the bulk region. The detailed information of atomic structure and chemistry is helpful to understand the electrical and ionic properties in these materials containing the dislocations and also provides valuable insights into prediction and design of dislocations and grain boundaries by controlling the sample treatment process.

**Results and Discussions**

Figure 1 (a) shows the schematic image of STO bicrystal with the tilt angle of 10°. Figure 1 (b) and (c) are the HAADF STEM images and the fast Fourier Transform (FFT) pattern, respectively. As shown in Fig. 1(b), the misorientation between the two single crystals leads to the formation of an array of dislocations. The FFT image shows two (100) diffraction spots that are rotated against each other by approximately 10°, which coincides with the designed tilt angle 10°. The distance measured between two dislocation cores is about 2.4 nm, which is in good agreement with the distance (2.2 nm) calculated by Frank's formula[30]:

$$d_{th} = \frac{|b|}{2\sin(\theta/2)}$$

where θ is the tilt angle (θ=10°) and │b│ the modulus of the Burgers vector (│b│ = 0.3905 nm). Figure 1(d) and (e) show the EDS maps of net counts for Sr and Ti, respectively, across the grain boundary. At the grain boundary, the net counts for Ti is reduced with dark color compared with that in the bulk. The line profile of net counts for Sr and Ti across the boundary in Fig. 1(f) indicates although both of Ti and Sr decrease at the boundary, reduction of Ti is much more significant than that of Sr, leading to the Ti deficiency dislocation cores at the grain boundary, which is different from the previous studies[28].

Furthermore, the atomically resolved EDS maps of Sr and Ti are carried out to estimate the composition of dislocations. **Fig.** 2 (a), (b) and (c) are net counts of Sr, Ti, and the intermix of Sr and Ti across the grain boundary. The corresponding intensity profiles of net counts for Ti and Sr elements across the grain boundary are shown in Fig. 2(d). From Fig. 2, two types of dislocation cores A and B appear at the grain boundary. At the dislocation core A and B, the brightness of Ti and Sr columns is very faint maybe due to the low occupancy. Besides, Sr atoms occupy Ti columns marked by the yellow circle in Fig. 2(b), leading to the Ti columns don't show pure green color in the mixed map in Fig. 2(c). From the line profile across the grain boundary, the net counts of Ti and Sr along the boundary are lower compared to the bulk regions, and the net counts of Ti decreased much more than that of Sr, same with the result mentioned in Fig. 1(f). From the Qmap, the ratio of Sr to Ti is 1:1.02 in the bulk, nearly 1:1 in perovskite structure, while it becomes 0.83:0.71 for dislocation core A, 0.88:0.67 for core B. The atomic-scale estimation of composition further confirms the phenomenon of Ti deficiency at the grain boundary.

**Figure** 3(a) shows the atomic resolution HAADF STEM image with two different dislocation cores. From Fig. 3 (a), it is clearly seen that the two types of dislocation cores A and B alternatively sit along the grain boundary. Since the intensity of atomic columns for the HAADF images is approximately proportional to the atomic number, the brighter spots in the image correspond to Sr-Sr columns, whereas the less bright ones are the Ti-O columns. Figure 3(b) shows an extracted HAADF-STEM image from the location marked by green rectangle in Fig. 3(a). And the termination plane for the dislocation core A is Ti-O plane, while the termination plane for dislocation core B is Sr-O plane, marked by yellow rectangles. Figure 3(c) shows the schematic structure containing core A and core B with the occupancy of individual Sr and Ti atomic columns, which can explain the Ti deficiency qualitatively. In bulk region, for one unite cell, the atomic ratio of Sr and Ti is 1:1, just like the crystal structure shown in Fig. 3(d). Within the trapezoid regions in Fig. 3(c), the occupancy of Ti columns is significantly decreased while more subtle

reduction occurs for Sr occupancy. For example, for core A, the Ti occupancy of for column1 is estimated to be 62% and for column 2 is 11% based on column intensity in the Z-contrast image compared with the bulk. And the occupancy of Ti for column 3 is about 100% which may partly result from the contribution from Sr atoms, since Sr substitution of Ti occurs at Ti column shown in Fig 2. For dislocation core B, the occupancy of Ti for columns 6 is 17% and for column 7 is 71%, while the Sr occupancy is about 78% and 83% for column 4 and column 5, respectively. As a result, under occupancy of Ti columns and Sr atoms partially occupies at the Ti columns the core regions both contribute the Ti deficiency.

Besides the elements distributions at the dislocation cores, the electronic structures of Ti are obtained from EELS measurement, which is effective to define the local chemical valence of Ti$^{32,33}$. The representative Ti spectra for the core A, core B, gap between A and B, and bulk matrix are shown in **Fig.4**. For the core A and B, the larger shift to low energy and the less peak splitting in the Ti L-edges are observed, corresponding to the low valence of Ti$^{28}$, while in the gap, the spectrum is almost the same with the bulk except slight shift. The relative intensity of Ti$^{3+}$ and Ti$^{4+}$ along the grain boundary is measured in Fig. 4 (c). At the dislocation cores A and B, density of Ti$^{3+}$ increases and Ti$^{4+}$ decreases. In fact, from the EELS measurements, significant reduction of O intensity is also observed in Fig. 4(d) and the reduction of O is even more than that of Ti, which explains the Ti$^{3+}$ can exist in the Ti deficient dislocation cores. It was reported before that the electronic and ionic transportation properties at the boundary is different from the bulk$^{4}$. The finding here of Ti deficient grain boundary in STO bicrystal also likely have important influence on its transportation properties and other physical properties, which needs further studies in future.

**Summary**

In summary, we study the atomic structure and chemistry of dislocations cores in 10° SrTiO$_3$ bicrystal that was fabricated at high temperature in N$_2$. The atomic arrangements and elemental occupancy are determined by Z-contrast imaging and atomically resolved EDS measurements. We find that the dislocation cores are Ti deficient, which is in contrast with previous studies of Ti richness at the dislocation cores. The Ti deficiency is mainly due to the under occupancy of Ti columns in the core regions. At the same time, some Sr atoms partially occupies at the Ti columns, which is another reason for the Ti deficiency. Moreover, the presence of O vacancy and reduced Ti from $Ti^{4+}$ to $Ti^{3+}$ are also confirmed at the dislocation cores from the EELS measurement. Our finding of Ti deficient dislocations suggests the atomic structure and chemistry of dislocations significantly depends on the fabrication process (i.e. heating treatment in $N_2$ in our study vs air in previous studies), providing valuable insights into understanding the relation between structure and properties for crystal defects. The ability to control the atomic structure and chemistry of dislocations/grain boundaries by bi-crystal fabrication also paves ways to design specific defects with unique properties.

**Methods**

**TEM sample preparation.** STO bicrystal with a [001]/(110) 10° tilt GB was fabricated by thermal diffusion bonding of two STO single crystals in the environment of $N_2$ protection. Thin foils for STEM observations were prepared by a conventional method that includes mechanical polishing and argon-ion beam milling. The ion-beam milling was carried out using PIPS$^{TM}$ (Model 691, Gatan Inc.) with the acceleration voltage of 3.5 kV until a hole was observed. Finally, low-voltage milling was carried out with the acceleration voltage of 0.5 kV to reduce the thickness of the irradiation-damaged layers.

**STEM observation and EDS analysis.** High-angle annular dark field (HADDF) images were recorded at 300 kV using an aberration-corrected FEI Titan Themis G2 with a spatial resolution up to ~60 pm. The convergence semi-angle for imaging is 15 mrad, and the collection semi-angles snap is 48 to 200 mrad. During imaging, the low electron doses were applied by using small aperture, small beam current and short scanning time. Typical HAADF images are shown in **Figure 1**b. To determine the occupancy of individual atomic columns in the dislocation cores, the intensity ratio of each atomic column in the cores to that in the bulk was calculated. We used the relation $I^{0.5}$ for HAADF images to estimate the occupancy, where I is the normalized intensity of the atomic columns in the dislocation cores. The EDS experiments were carried out at 300 kV. Typical net count EDS maps for Sr, Ti and O are shown in **Figure 1**b-c and **Figure 2**a-c. To obtain the atomic percentage of Sr and Ti, the quantitative map (Qmap) was conducted. The same pixels were selected to obtain the ratio of Sr and Ti.

**EELS analysis.** The electron energy loss spectroscopy (EELS) experiments were carried out at 300 kV. The electron beam was slightly spread and the acquisition time is 0.5 s/pixel. Spectra of line scan from 400~490 eV is obtained with energy dispersion 0.025 eV. And EELS mapping with region of ~7×3.5 nm$^2$ from 350~860 eV is obtained with energy dispersion 0.1eV, shown in **Figure 4** and **Figure S3**. The distribution of Ti$^{3+}$ and Ti$^{4+}$ was obtained by MLLS fitting and quantification with the software of Digital Microscopy (Gatan).

**Acknowledgement**

This work was supported by National Equipment Program of China (ZDYZ2015-1), the National Key R&D Program of China (2016YFA0300804, 2016YFA0300903), National Natural Science Foundation of China (51672007, 51502007, 11474337, and 51421002), the National Program for Thousand Young Talents of China and "2011 Program" Peking-Tsinghua-IOP Collaborative Innovation Center of Quantum Matter. The authors also acknowledge Electron Microscopy Laboratory in Peking University for the use of Cs corrected electron microscope.

**Author contributions**

X.J. Li made the specimen and perfromed on STEM with the assistance of X.M. Ma and J.M. Zhang. X.J. Li, S.L. Chen, and M.Q. Li analyzed the results under the guidance of X.D Bai, P. Gao and K.H. Liu, and finished the manuscript.

**Additional information**

Supplementary information
Competing Interests: The authors declare no competing interests.

**Figures and captions**

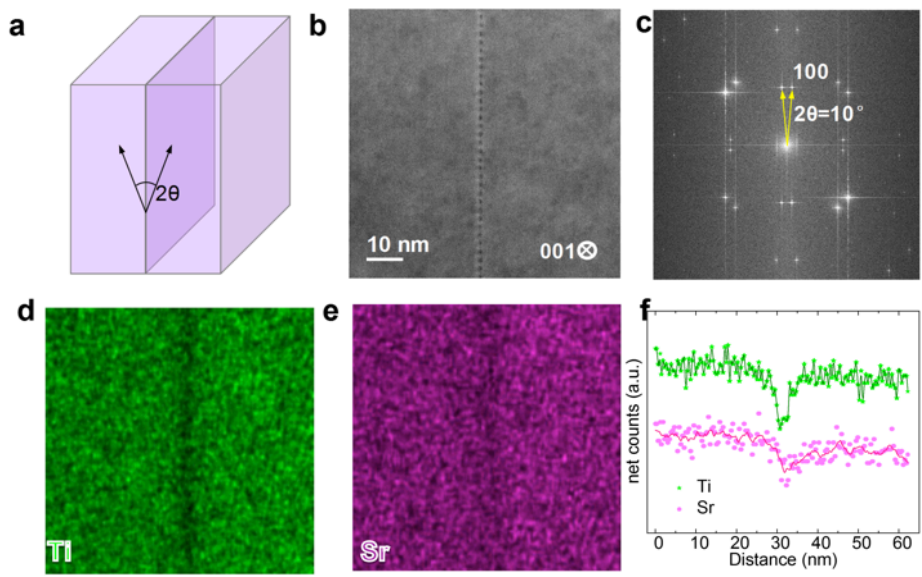

**Figure 1.** (a) Schematic illustration of the fabricated STO bicrystal with the tilt angle of 10°. (b) A HAADF-STEM image, and (c) the corresponding Fast Fourier Transform (FFT) confirming the tilt angle is 10°. (d) Energy dispersive X-ray spectra (EDX) with the net count map of Sr. (e) EDS with the net count map of Ti. (f) Line profiles of Sr and Ti across the grain boundary.

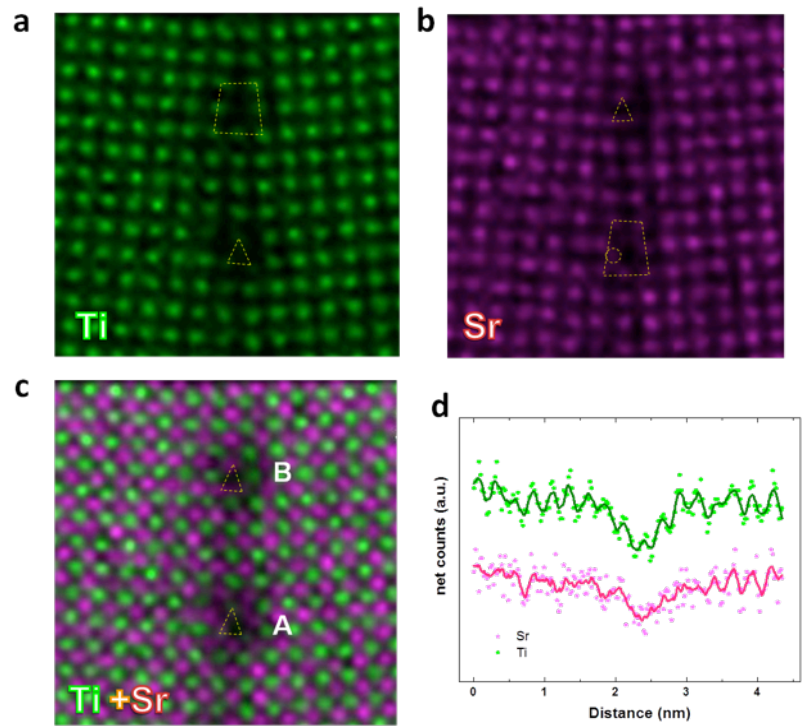

**Figure 2.** The atomically resolved EDS maps around the dislocations. (a) Net count map of Sr. (b) Net count map of Ti. (c) Intermix of Sr and Ti. (d) Line profiles of Ti and Sr across the dislocation cores.

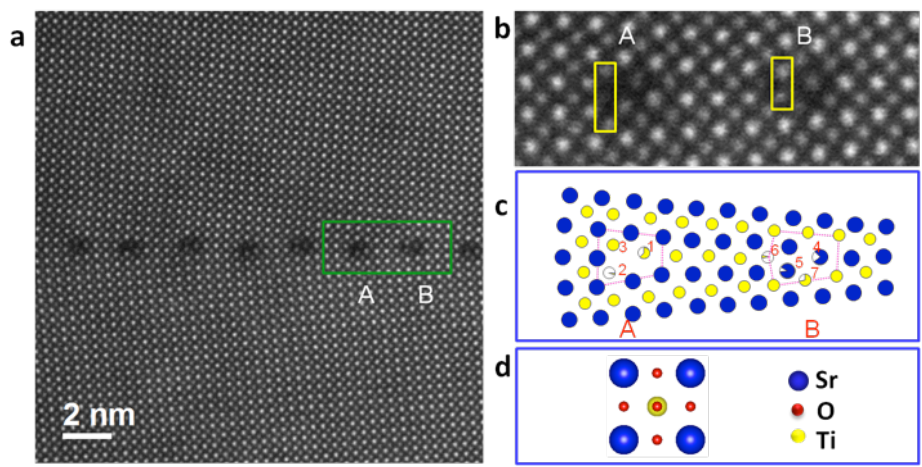

**Figure 3.** (a) Atomic resolution HAADF-STEM image. (b) An extracted HAADF-STEM image from the location marked by the green rectangle in (a). (c) Schematic core-structure with the sector indicating the possible occupancy. (d) Crystal structure of SrTiO3 viewing from [001] direction.

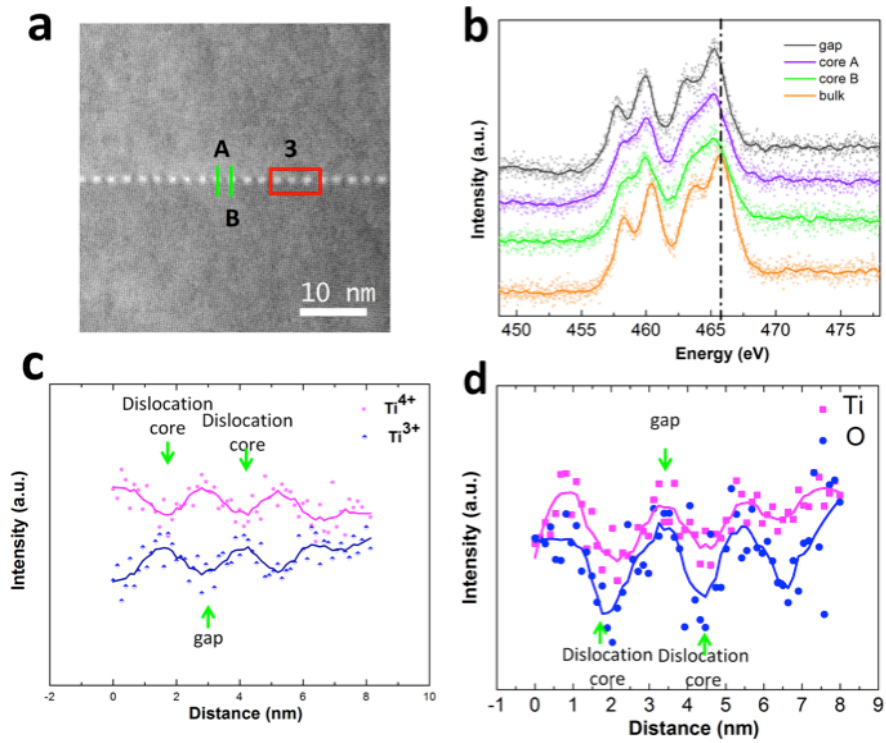

**Figure 4.** (a) The position of EELS line scans (green line1 and 2) and mapping (red rectangle 3). (b) EEL spectral of core A, core B, gap between core A and core B, and bulk aside the boundary. (c) Distribution of $Ti^{3+}$ (blue) and $Ti^{4+}$ (pink) from EELS mapping result by MLLS fitting. (d) Distribution of Ti (blue) and O (pink) from EELS mapping result by quantification.